\documentclass[conference,a4paper]{IEEEtran}
\IEEEoverridecommandlockouts
\usepackage{cite}
\usepackage{amsmath,amssymb,amsfonts}
\usepackage{textcomp}
\usepackage{xcolor}
\usepackage{multirow}
\usepackage{nccmath}
\usepackage{mathtools}
\usepackage{graphicx}
\usepackage[strings]{underscore}
\usepackage{xcolor,colortbl}
\usepackage[ancient]{flushend}
\usepackage{upgreek}
\usepackage{caption}
\usepackage{subcaption}

\def\BibTeX{{\rm B\kern-.05em{\sc i\kern-.025em b}\kern-.08em
    T\kern-.1667em\lower.7ex\hbox{E}\kern-.125emX}}

\graphicspath{ {./assets/} }

\begin{document}

\title{Maximum Power Point Tracking Circuit for an Energy Harvester in 130 nm CMOS Technology}

\author{
	\IEEEauthorblockN{
		Adam Hudec, Lukas Nagy, Martin Kovac and Viera Stopjakova
	}
	\IEEEauthorblockA{
		\textit{Institute of Electronics and Photonics} \\
		\textit{Faculty of Electrical Engineering and Information Technology} \\
		\textit{Slovak University of Technology} \\
		Bratislava, Slovakia \\
	}
}
\maketitle

\begin{abstract}
This paper presents design of a Maximum Power Point Tracking (MPPT) circuit and its functionality for tuning the maximum power transfer from an energy harvester (EH) unit. Simple and practical ``Perturb and Observe'' algorithm is investigated and implemented. We describe the circuit functionality and the improvements that have been introduced to the original algorithm. The proposed MPPT design is divided into three main blocks. The output signal is being generated by the PWM or PFM block. The tracking speed has been enhanced by implementing a variable step size in the ``Tracking Block''. Finally, the overall power consumption of the MPPT circuit itself is controlled by the ``Power Management Block'', which manages delivering the clock signal to the rest of the circuit. The RTL code of the proposed MPPT has been described in Verilog, then has been synthesized and placed-and-routed in a general purpose 130nm CMOS technology.   
\end{abstract}

\section{Introduction}
Once upon a time, the electronics were not developed or it was underdeveloped, so people thought about what and how to make from the mechanical parts. However, recent technologies and possibilities enable development of better and better electronic devices. Nowadays, we experience a major boom in Internet of Things (IoT). We try to connect well-nigh all the electronics to the Internet so that we can monitor or control it remotely. We look for free parking places in towns via smartphones, monitor car tire pressure via on-board computer, control household appliances with voice and etc. We just want to have everything SMART. During the development, more and more emphasis is placed on the functionality at lower supply voltage and reducing their total current consumption. That is why many portable or wearable devices need to be battery-powered, only. Energy harvesters can also help to prolong the battery life of given electronic systems. EHs are increasinghly becoming an integral part of many electronic applications, mainly the sensor systems \cite{salazar}. The stand-alone harvesters that work without the support and management circuit, rarely deliver the output power with the maximum efficiency. Furthermore, the enviromental conditions of the EH will inevitebly vary over time, so a management block, which fine-tunes its operating point accordingly, is crucial. The vast majority of MPPT circuits is based on a digital control of the energy harvester implemented in a feed-back loop. One of the most widely used algorithms for MPPT is called ``Perturb and Observe'' \cite{banu, 2, 3}. As the name suggests, it perturbs a given parameter of a specific EH signal (e. g. frequency, duty cycle, amplitude, etc.) and compares the actual power transfer value with the previous one. These differences may vary in size, therefore we have implemented the variable step size for faster tuning of the maximum power point. Because EH helps extend battery life, there is no need to operate at the highest clock frequency even after converging to the current total efficiency. In order to ensure the lowest quiescent power consumption, the circuit operates with variable clock frequency depending on the change of input power magnitude of the energy harvester.
	
\section{Maximum Power Point Tracking}
\begin{figure}[!h]
	\begin{center}
		\includegraphics[width=0.49\textwidth]{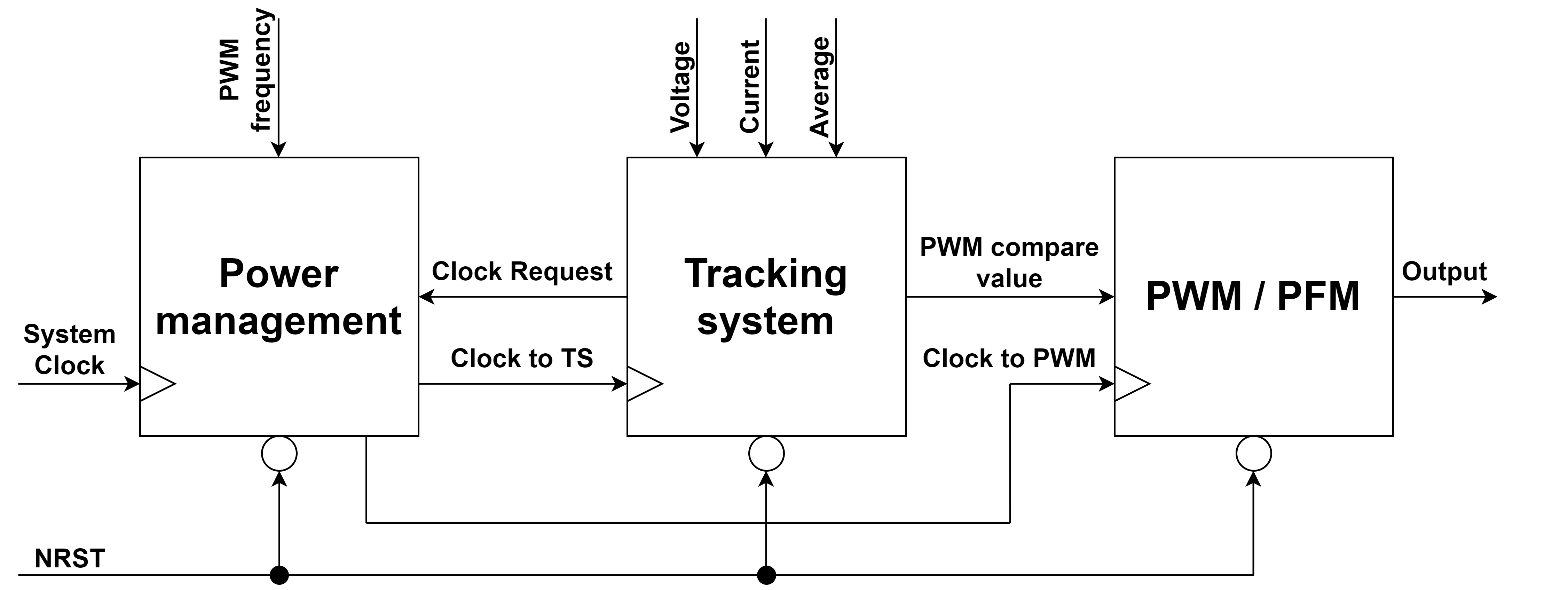}
	\end{center}
	\caption{Block schematic diagram of the proposed MPPT.}
	\label{fig:MPPT}
\end{figure}

The MPPT algorithm ensures the optimal usage of energy harvesters in general. \cite{bizon2017energy, kazmierski2014energy, mohapatra}. For example, it helps the DC-DC converter to increase or decrease the output voltage based on the conditions of the energy source, in order to maintaining the power transfer efficiency. This adjustment is achieved by tuning the PWM / PFM signal used for switching the power transistors within the DC-DC converter. A feed-back loop is used, so the power transfer is always maximized under any operating conditions. In Fig. \ref{fig:MPPT}, we can see the block diagram of the proposed digital MPPT design with added improvements in the power management block. Our entire design is described in Verilog at Register-Transfer Level (RTL).

\subsection{Tracking System}
\begin{figure}[!h]
	\begin{center}
		\includegraphics[width=0.49\textwidth]{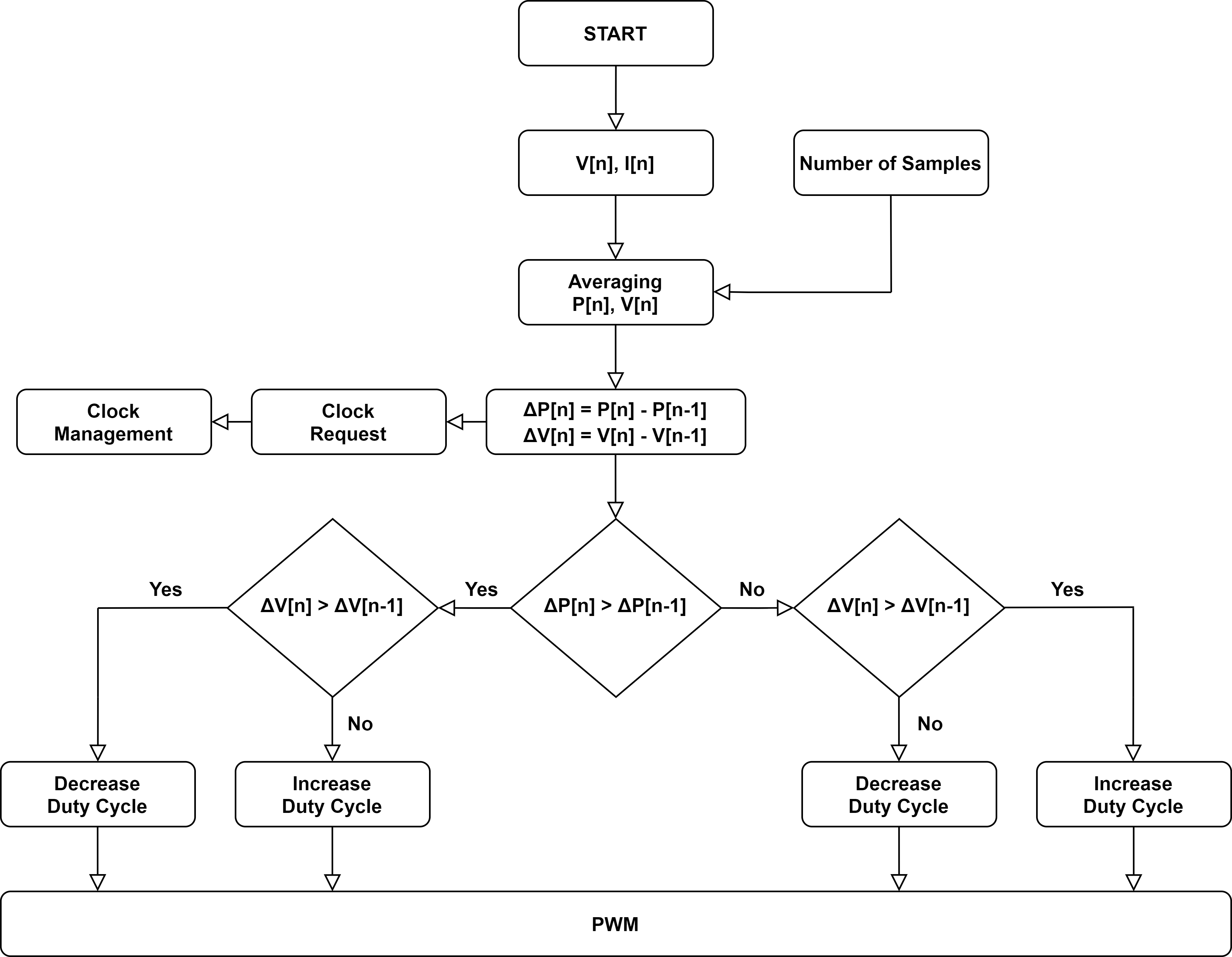}
	\end{center}
	\caption{Flow Chart of Tracking System.}
	\label{fig:TRACKER}
\end{figure}

	As mentioned before, the MPPT tracking system is based on the Perturb and Observe method \cite{john2017variable, 4}. It is a direct algorithm that measures the generated voltage and current in the real time, or in other words, the instant power transfer. The Tracking System, which is a part of the block diagram in Fig. \ref{fig:MPPT}, has two outputs and five inputs including the clock and global reset signal. Inputs like voltage and current are 8-bit numbers, from which the instant power is calculated. Due to expected fluctuation of the input values, we implemented their averaging, which in return improves the stability and accuracy of the feed-back loop. The number of samples involved in the averaging process is determined externally through a 2-bit input called "Averaging". The states on this 2-bit bus do not directly refer to the exact number of samples, but they are a custom form of code that conceals the exact number of averaging samples. In our propose we work with the number of samples 1, 8, 32 and 64. After the average value is calculated for the input values, the actual current and voltage values are compared with their previous values. In Fig. \ref{fig:TRACKER}, one can observe, that after this step there are two different possible events, which are executed in the same time. In the first case, the duty cycle of the output PWM signal increases or decreases based on the evaluation of the input conditions. The resolution of the duty cycle step size is determined by the 8-bit output called "PWM compare value". In the second case, the input frequency of the Tracking System block is changing based on the magnitude of the instant power change. More information and deeper description are discussed in the following subsection of the paper.

\subsection{Power Management} \label{PWRMNG}
\begin{figure}[!h]
	\begin{center}
		\includegraphics[width=0.25\textwidth]{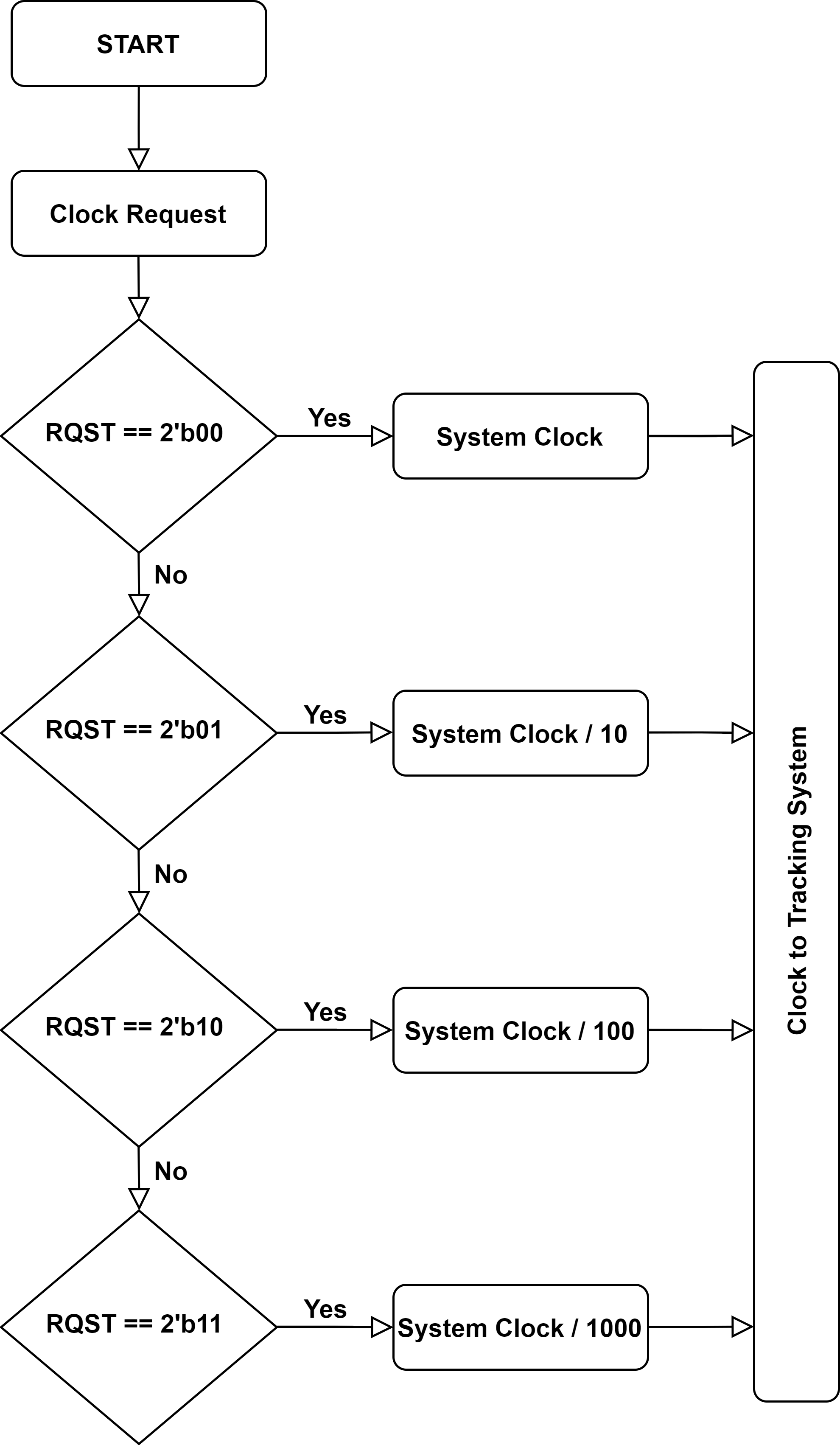}
	\end{center}
	\caption{Flow Chart of Power Management.}
	\label{fig:PWRMNG}
\end{figure}

	The proposed MPPT block is a digital synchronous circuit that needs a clock signal to function. In our case, it is called "System Clock", which enters into the Power Management. It is based on the Request and Response method and frequency dividers. In order to reduce the MPPT internal power consumption as much as possible, we created a module that manages the distribution of the variable frequency clock signal to the Tracking System and to the PWM/PFM block. As shown in Eq. \ref{eq:POWER}, one way to reduce the dynamic power consumption of a digital circuit is to reduce its operating frequency.
\begin{equation}
	P_{dynamic} = \alpha C_L V^2 f_{clk} 
	\label{eq:POWER}
\end{equation}
The first step to reduce the power consumption is externally set the output PWM frequency. It is possible to use an 8-bit input value, called "PWM frequency". The second, and for us one of the most effective options to reduce the power consumption, is the dynamic change of clock frequency for the Tracking System module. Since it constantly compares the actual value of the power with its previous value, it is possible to determine the change of working frequency. If the change of detected power transfer is low or close to zero, the MPPT has tuned the energy harvester to the maximum power point and therefore there is no need for delivering a high clock frequency to the Tracking System block. It itself evaluates and sends requests to the Power Management module to keep or change the clock frequency as shown in Fig. \ref{fig:PWRMNG}. These requests are sent as a 2-bit number via bus called "Clock Request". As with the Tracking System, it is not a direct code but a custom interpretation of a request. Power Management evaluates these requests and then distributes the System Clock signal via divider to the Tracking System. Frequency dividers are sequential circuits made of counters. These are sensitive to the rising edge of System Clock and therefore the highest frequency we can supply to the Tracking System is a half of System Clock. The dividing ratio which we have chosen varies by the decimal value and, due to the above-mentioned property of the sequential circuit also by a multiple of 2. The resulting split ratios are thus 2, 20, 200 and 2000.

\subsection{Pulse Width or Pulse Frequency Modulation}
\begin{figure}[!h]
	\begin{center}
		\includegraphics[width=0.18\textwidth]{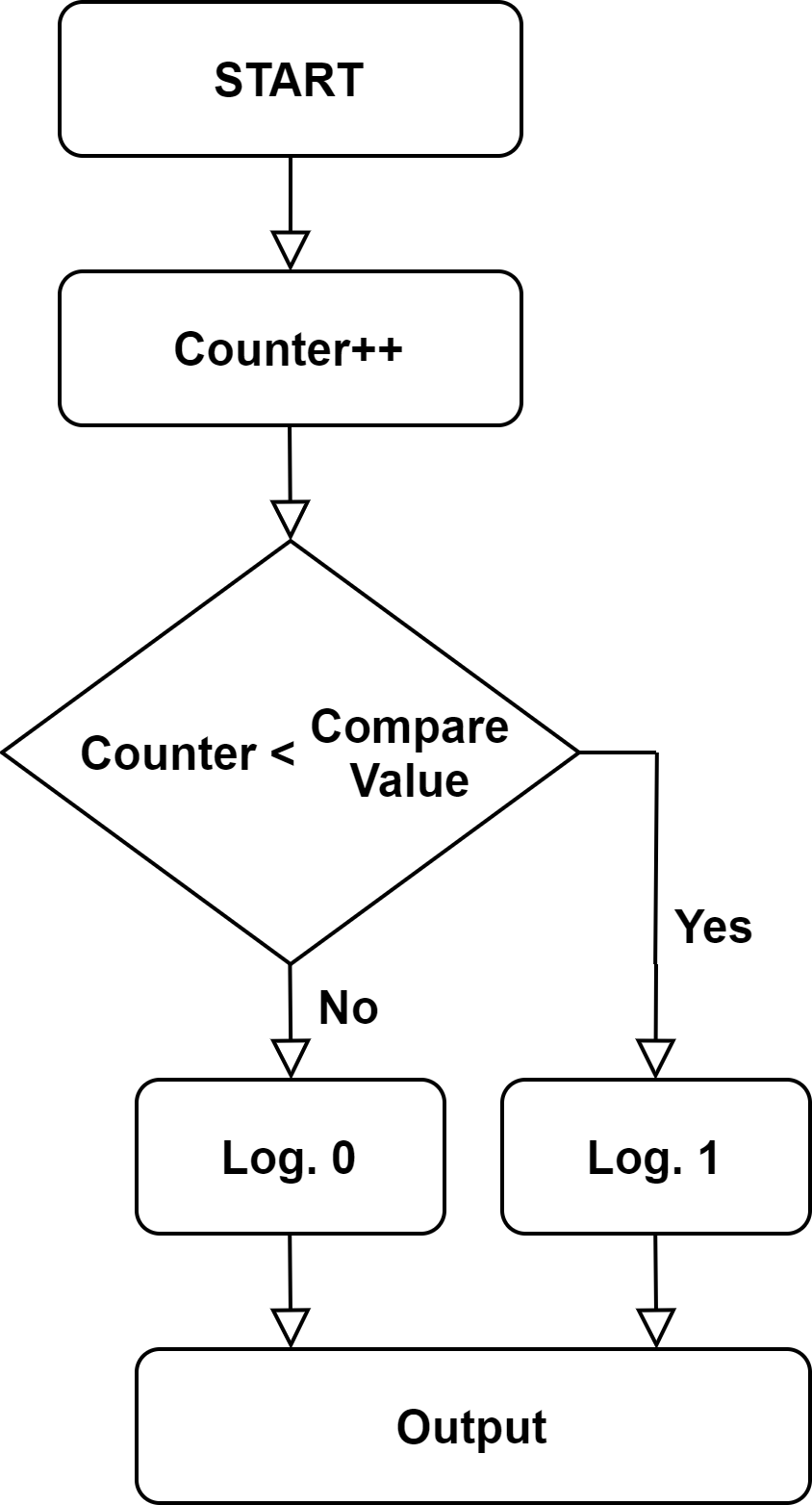}
	\end{center}
	\caption{Flow Chart of PWM}
	\label{fig:PWM}
\end{figure}

	The MPPT output stage is designed to be modular. For a DC-DC converter, it is possible to implement Pulse Width Modulation (PWM) or Pulse Frequency Modulation (PFM) to drive the power transistors switching the inductor or transformer \cite{carvalho2015cmos}. In our design, PWM generator is primarily used. Its output signal is rectangular with constant frequency, but with variable \emph{duty cycle}, which is defined by equation Eq. \ref{eq:RATIO}.
\begin{equation}
	DC = \left(\frac{T_{Log. 1}}{T_{Log. 1} + T_{Log. 0}}\right) * 100 \%
	\label{eq:RATIO}
\end{equation}
In the previous subsection \ref{PWRMNG}, the setting of the operating frequency of the PWM block was mentioned. As one can see, in Fig. \ref{fig:MPPT}, an 8-bit bus called "PWM frequency" is used for a static PWM frequency setting. The design itself consists of a comparator and an 8-bit counter that increments from 0 to 255 by 1 at the clock rising edge. This ensures a constant signal period. The duty cycle is based on the compared value which is coming from the Tracking System depending on the power changes on the MPPT input. From the flowchart in Fig. \ref{fig:PWM} one can see that if the counter value is less than the compared value, on the output there is logic 1, otherwise logic 0 is asserted. Moreover, during incrementing until the counter is not in the overflow state, it is not possible to update the compare value from the Tracker System in order to avoid changing the frequency. Default state of the duty cycle is set to 50 \%. This setting is based on the assumption that the Maximum Power Point is situated around the center of the ideal curve, so in the real case, the tuning will take considerably less time.

\section{Post-Layout Simulation Results}
A complete RTL design has been synthesized and placed-and-routed in 130 nm CMOS technology using Cadence tools. The total number of standard cells is 743 including the clock tree and physical cells with 10 234~$\mu$m$^{2}$ of occupied area and totalling with 10 436 transistors. The finished physical design has been afterwards imported into the \emph{analog} environment and simulated in various PVT conditions with clock signal of $f_{clk}$ = 1~MHz. The main constrain, in our case, was already mentioned power consumption. A very effective way of reducing the power consumption is to decrease the supply voltage $V_{DD}$. The system remained fully functional with $V_{DD}$ = 0.4~V in all process and temperature corners. Table \ref{tab:P} summarizes the simulated average power consumption of the proposed MPPT design. As expected, the highest demand for power occurs with fast transistors and high temperature. However, the decrease in power consumption with $V_{DD}$ = 0.4~V is immense. The value has been determined at 92.71~\%, which is truly astonishing.

\begin{figure}[!h]
	\begin{center}
		\includegraphics[width=0.49\textwidth]{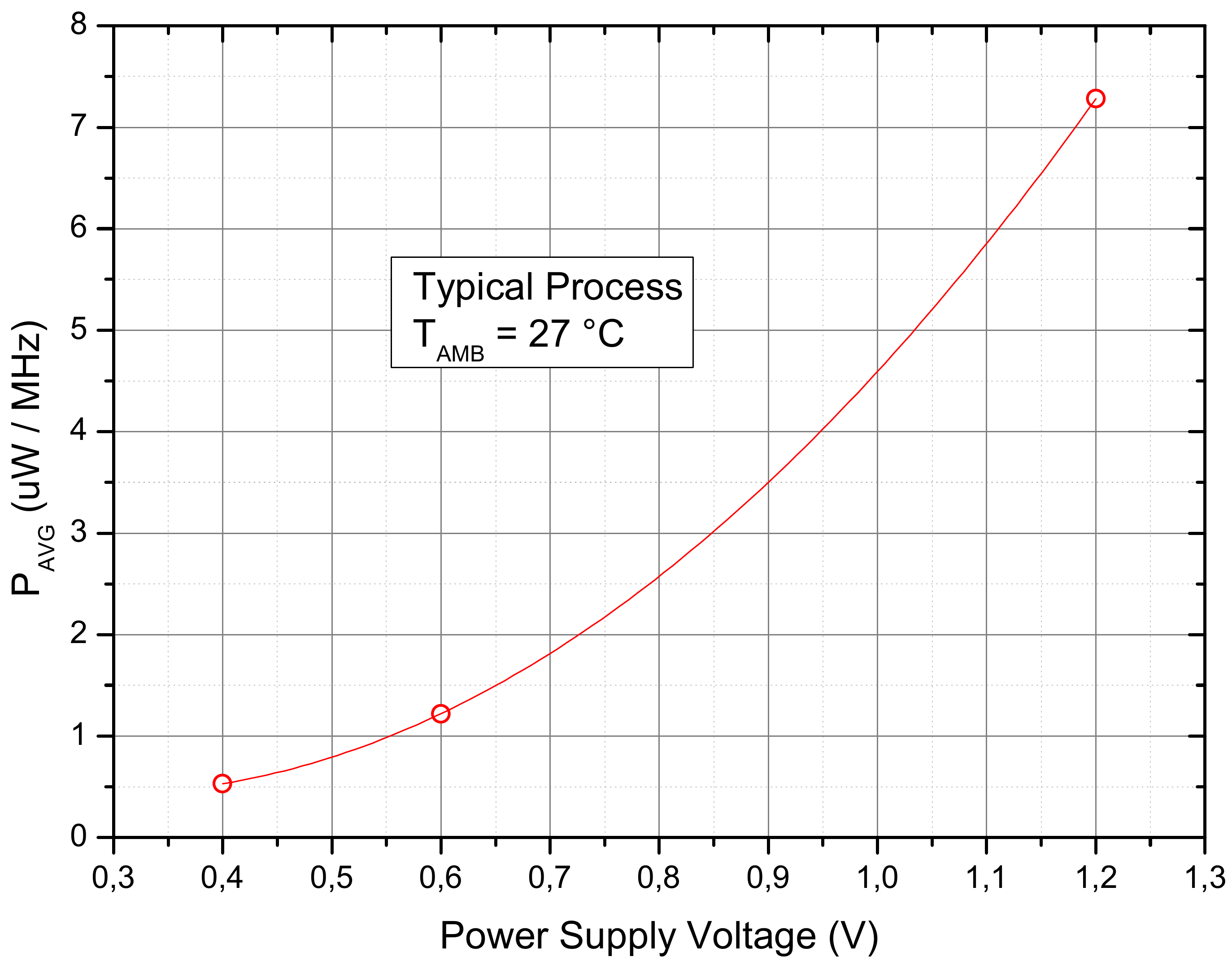}
	\end{center}
	\caption{Power consumption as a function of power supply voltage at nominal conditions.}
	\label{fig:graph1}
\end{figure}

\begin{figure}[!h]
	\begin{center}
		\includegraphics[width=0.49\textwidth]{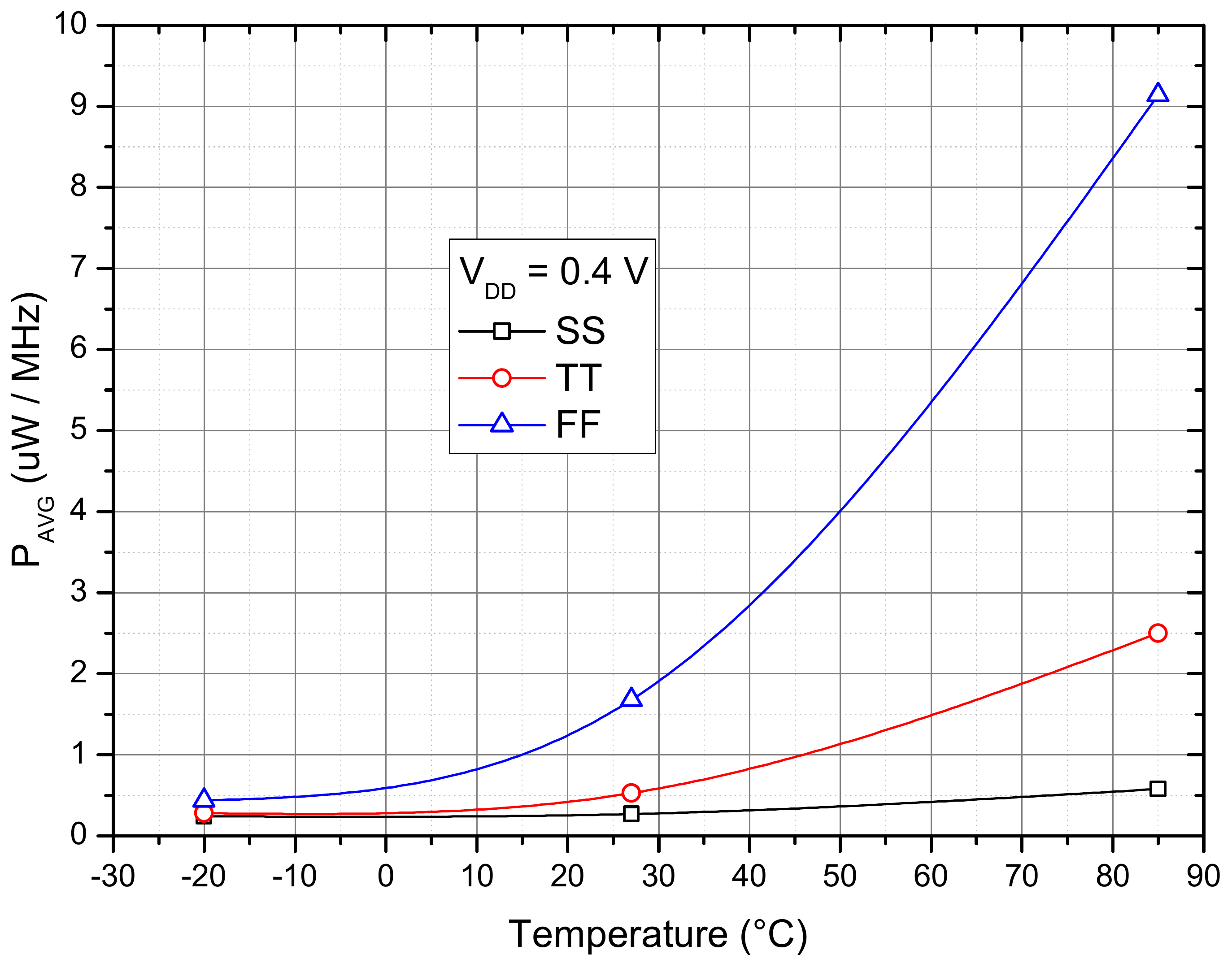}
	\end{center}
	\caption{Power consumption as a function of temperature for lowered supply voltage.}
	\label{fig:graph2}
\end{figure}

\begin{table*}[]
\caption{Average power consumption of the proposed MPPT in various PVT conditions.}
\centering
\begin{tabular}{|c|c|c|c|c|c|c|c|c|c|}
\hline
$P_{avg}$ (uW/MHz)    & \multicolumn{3}{c|}{$V_{DD}$ = 1.2 V} & \multicolumn{3}{c|}{$V_{DD}$ = 0.6 V} & \multicolumn{3}{c|}{$V_{DD}$ = 0.4 V} \\ \hline
Process Corner	      & SS  	    	& TT      	& FF        	& SS         	& TT         	& FF         	& SS        	& TT         	& FF   	 \\ \hline
T = -20 $^\circ$C     & 2.79 		& 4.15 		& 7.87    	& 0.56  	& 0.73		& 1.05  	& 0.24 		& 0.28  	& 0.44    \\ \hline
T = 27 $^\circ$C      & 3.29  		& 7.28		& 24.3  	& 0.64   	& 1.22   	& 3.78   	& 0.27   	& 0.53   	& 1.68    \\ \hline
T = 85 $^\circ$C      & 6.42     	& 27.4 		& 100 		& 1.28   	& 5.26   	& 19.1   	& 0.58   	& 2.5    	& 9.14 	 \\ \hline
\end{tabular}
\label{tab:P}
\end{table*}
The Fig. \ref{fig:graph1} depicts the dependence of power consumption a a function of supply voltage at typical process and ambient temperature. The dependence is expected to be quadratic (Eq. \ref{eq:POWER}). The lowest power supply voltage the system could tolerate in all process corners was found to be $V_{DD}$ = 0.4~V. Therefore, the deeper investigation has been carried out with mentioned supply voltage level.
Fig. \ref{fig:graph2} shows the dependence of average power consumption with $V_{DD}$ = 0.4~V as a function of ambient temperature for the worst, best and nominal fabrication process corners. The dependence is exponential and varies over wide range of values, as expected. However, the average power consumption never exceeds 10~uW / MHz and in the nominal process corner, the power consumption remains firmly in the range of nW, which is truly astonishing result for such an old fabrication process.
%

\section{Conclusion}
We presented the design of the MPPT circuit using Perturb and Observe algorithm, which is a part of the energy harvesting system. The emphasis of the design was focused on the non-functional parameter, namely the total power consumption. As mentioned above, one of the ways to significantly reduce power consumption is switch the operating frequency of the circuit to a lower value. We succeeded in the improvement of the Perturb and Observe method, where a variable convergence step based on a large variability of differences in the input power over time, was implemented. This means that tuning the Maximum Power Point takes less time than tuning with a constant step. This is closely linked to our added Power Management enhancement. It changes and distributes the working frequency to the Tracker System, which after the convergence to the MPP sends a request to reduce the clock frequency to the minimum. The proposed design has been successfully synthesized and placed-and-routed into 130 nm CMOS technology. The average power consumption (leakage and dynamic combined) was simulated in all process and temperature corners with three different supply voltages. The worst-case power consumption with $V_{DD}$ = 0.4~V is $P_{avg}$ = 8.8~$\mu$W/MHz, while in typical process and temperature conditions, it falls safely into the nW range. Namely, $P_{avg}$ = 0.528~$\mu$W/MHz, which represents reduction of astonishing 92.7 \% from the nominal value. One of the improvements we want to address next, is to further reduce dynamic circuit power consumption by adding more clock gating points and finding a trade-off between the power consumption and convergence rate to the Maximum Power Point. Another improvement to the MPPT circuit which will also increase the Energy Harvester efficiency in degraded conditions, is the dynamic switching of the MPPT output block between PWM and PFM output signal to DC-DC converter. 
\\
\section{Acknowledgement}
This work was supported in part by the Ministry of Education, Science, Research and Sport of the Slovak Republic under grants VEGA 1/0905/17 and VEGA 1/0731/20, and ECSEL JU under project PROGRESSUS (876868).

\nocite{*}
\bibliographystyle{IEEEtran}
\bibliography{references}
\end{document}